\documentstyle[amssymb,12pt,epsfig]{article}

\def\beq{\begin{equation}}
\def\eeq{\end{equation}}
\newcommand{\newc}{\newcommand}
 \newc{\sm}{Standard Model }
 \newc{\smd}{Standard Model}
 \newc{\dac}{discrete anomaly cancellation }
 \newc{\mup}{``$\mu$'' problem }
\newcommand{\be}{\begin{equation}}
\newcommand{\ee}{\end{equation}}
\newcommand{\bea}{\begin{eqnarray}}
\newcommand{\eea}{\end{eqnarray}}

\def\be{\bar{\epsilon}}

\def\be{\bar{\epsilon}}

 \newc{\eps}{\epsilon}
 \newc{\barr}{\begin{eqnarray}}
 \newc{\earr}{\end{eqnarray}}
\def\gappeq{\mathrel{\rlap {\raise.5ex\hbox{$>$}}
 {\lower.5ex\hbox{$\sim$}}}}
\def\lappeq{\mathrel{\rlap{\raise.5ex\hbox{$<$}}
 {\lower.5ex\hbox{$\sim$}}}}

\begin{document}

\title{\begin{flushright}
\vspace*{-1.2 cm}
{\normalsize hep-ph/9902283} \\
\vspace*{-0.4 cm}
{\normalsize CERN-TH/99-27}  \\
\vspace*{-0.4 cm}
{\normalsize OUTP-99-11P} 
\end{flushright} 
\vspace*{0.7 cm} {\large {\bf {\ }}}{\bf Neutrino masses from } %
\mbox{\boldmath{$U(1)$}}{\bf \ symmetries}\\
{\bf \ \ \ \ \ \ and the Super-Kamiokande data }\\
}
\author{{\ {\bf Smaragda Lola$^{a}$ and Graham G. Ross$^{b}$}} \\
\begin{tabular}{l}
$^{a}$ {\small CERN Theory Division, Geneva 23, CH-1211, Switzerland } \\ 
$^{b}${\small Department of Physics, University of Oxford, OX1 3NP, UK}
\end{tabular}
}
\date{}
\maketitle

\begin{abstract}
{\small {Motivated by the Super-Kamiokande data, we revisit models with $%
U(1) $ symmetries and discuss the origin of neutrino masses and mixings in
such theories. We show that, in models with just three light neutrinos and a
hierarchy of neutrino masses, large (2-3) mixing fixes the lepton doublet }}$%
U(1)$ {\small {charges and is thus related to the structure of the charged
lepton mass matrix. We discuss the fermion mass structure that follows from
the Abelian family symmetry with an extended gauge group. Requiring that the
quark and lepton masses be ordered by the family symmetry, we identify the
most promising scheme. This requires large, but not necessarily maximal,
mixing in the }}$\mu \tau ${\small \ sector and gives }$e\mu $ {\small mixing%
} {\small {in the range that is required for the small angle solution of the
solar neutrino deficit. }}
\end{abstract}

\section{Introduction\label{sec:neutrinomass}}

\bigskip {\small Recent reports by the Super-Kamiokande collaboration~\cite
{SKam}, indicate that the number of $\nu _{\mu }$ in the atmosphere is
decreasing, due to neutrino oscillations. These reports seem to be supported
by the recent findings of other experiments \cite{KamMac}, as well as by
previous observations \cite{Hir}. The data indicates that the number of $\nu
_{\mu }$ is almost half of the expected number, while the number of $\nu
_{e} $ is consistent with the expectations. $\nu _{\mu }-\nu _{\tau }$
oscillations, with 
\begin{eqnarray}
\delta m_{\nu _{\mu }\nu _{\tau }}^{2} &\approx &(10^{-2}\;{\rm to}%
\;10^{-3})\;{\rm eV^{2}} \\
sin^{2}2\theta _{\mu \tau } &\geq &0.8  \label{atmos}
\end{eqnarray}
match the data very well, while dominant $\nu _{\mu }\rightarrow \nu _{e}$
oscillations are disfavoured by Super-Kamiokande~\cite{SKam} and CHOOZ ~\cite
{chooz}. }

{\small On the other hand, the solar neutrino puzzle can be resolved through
matter enhanced oscillations \cite{MSW} with either a small mixing angle: 
\begin{eqnarray}
\delta m_{\nu _{e}\nu _{\alpha }}^{2} &\approx &(3-10)\times 10^{-6}\;{\rm %
eV^{2}}  \label{smallmsw} \\
sin^{2}2\theta _{\alpha e} &\approx &(0.4-1.3)\times 10^{-2}
\end{eqnarray}
or a large mixing angle: 
\begin{eqnarray}
\delta m_{\nu _{e}\nu _{\alpha }}^{2} &\approx &(1-20)\times 10^{-5}\;{\rm %
eV^{2}} \\
sin^{2}2\theta _{\alpha e} &\approx &(0.5-0.9)
\end{eqnarray}
or vacuum oscillations: 
\begin{eqnarray}
\delta m_{\nu _{e}\nu _{\alpha }}^{2} &\approx &(0.5-1.1)\times 10^{-10}\;%
{\rm eV^{2}} \\
sin^{2}2\theta _{\alpha e} &\geq &0.67
\end{eqnarray}
where $\alpha $ is $\mu $ or $\tau $\footnote{{\small Best fit regions for
solutions to the solar neutrino deficit have been identified in \cite{BAH} .}%
}. }

{\small If neutrinos were to provide a hot dark matter component, then the
heavier neutrino(s) should have mass in the range $\sim (1-6)$ eV, where the
precise value depends on the number of neutrinos that have masses of this
order of magnitude \cite{mixed}. Of course, this requirement is not as
acute, since there are many alternative ways to reproduce the observed
scaling of the density fluctuations in the universe. }

{\small Finally, let us note that there is another indication of neutrino
mass. The collaboration using the Liquid Scintillator Neutrino Detector at
Los Alamos (LSND) has reported evidence for the appearance of $\bar{\nu}%
_{\mu }-\bar{\nu}_{e}$ \cite{LSND1} and ${\nu }_{\mu }-{\nu }_{e}$
oscillations \cite{LSND2}. Interpretation of the LSND data favours the
choice 
\begin{eqnarray}
0.2\,{\rm eV}^{2} &\leq &\delta m^{2} ~~~~\leq~ 10 \, {\rm eV}^{2}  \nonumber
\\
0.002 &\leq &\sin ^{2}2\theta ~\leq~ 0.03
\end{eqnarray}
The experiment KARMEN 2 \cite{KARMEN2} (the second accelerator experiment at
medium energies) is also sensitive to this region of parameter space and
restricts the allowed values to a relatively small subset of the above
region. }

The implications of these measurements are very exciting, for non-zero
neutrino mass means a departure from the Standard Model and neutrino
oscillations indicate violation of lepton family number, again lying beyond
the Standard Model. The first question that needs to be answered, is why are
the neutrino masses so small. In this paper we will follow what we believe
to be the most promising explanation, namely that neutrino masses are small
due to the ``see-saw'' mechanism \cite{seesaw} in which the light neutrino
are suppressed by a very large scale associated with the onset of new
(unified?) physics. The ``see-saw'' mechanism follows naturally in the case
that right-handed neutrinos exist.

Suppose that there is no weak isospin 1 Higgs field and hence there are no
mass terms of the $\nu_L \nu_L$ type. In this case there are two possible
neutrino masses

\begin{equation}
m_{Dirac}\overline{\nu _{L}}\nu _{R}+M_{Majorana}\nu _{R}\nu _{R}\equiv
\left( 
\begin{array}{cc}
\overline{\nu _{L}} & \nu _{R}
\end{array}
\right) \left( 
\begin{array}{cc}
0 & m_{D} \\ 
m_{D} & M_{M}
\end{array}
\right) \left( 
\begin{array}{c}
\overline{\nu _{L}} \\ 
\nu _{R}
\end{array}
\right)  \label{seesaw}
\end{equation}
The Dirac mass is similar to an up quark mass and one's naive expectation is
that they should have similar magnitude. On the other hand, the Majorana
mass term is invariant under the Standard Model gauge group and does not
require a stage of electroweak breaking to generate it. For this reason, one
expects the Majorana mass to be much larger than the electroweak breaking
scale, perhaps as large as the scale of the new physics beyond the Standard
Model; for example, the Grand Unified scale or even the Planck scale.
Diagonalising the mass matrix gives the eigenvalues 
\begin{eqnarray}
m_{Heavy} &\simeq &M_{M}  \label{eigenvalues} \\
m_{Light} &\simeq &\frac{m_{D}^{2}}{M_{M}}  \nonumber
\end{eqnarray}
The see-saw mechanism generates an effective Majorana mass for the light
neutrino (predominantly $\nu _{L}$) by mixing with the heavy state
(predominantly $\nu _{R})$ of mass $M_{M}.$ It is driven by an effective
Higgs $\Phi ^{I_{W}=1}$ made up of $H^{I_{W}=1/2}H^{I_{W}=1/2}/M_{M}$ (hence
the two factors of $m_{D}$ in eq.(\ref{eigenvalues})$~).$ eq.(\ref
{eigenvalues}) shows that a large scale for the Majorana mass gives a very
light neutrino. For example with $M_{M}=10^{16}~{\rm GeV}$ and $m_{D}$ taken
to be the top quark mass gives 
\[
m_{Light} ~\simeq~ 3.10^{-3} ~{\rm eV } 
\]

This estimate shows that it is quite natural to have neutrinos in a mass
range appropriate to give, for example, solar neutrino oscillations.
However, in many cases, larger masses capable of explaining the other
oscillation phenomena are possible because the Majorana mass for the right
handed neutrinos is often smaller than the Grand Unified mass. A Majorana
mass for the right-handed neutrino requires a Higgs carrying right-handed
isospin 1 (in analogy with the left-handed case when it needed left-handed
isospin 1). If this field is not present (for example in level one string
theory this is always the case) one may get a double see-saw because the
Majorana mass for the right-handed neutrino is also generated by an
effective Higgs, made up of $H^{I_{W,R}=1/2}H^{I_{W,R}=1/2}/M^{{\prime }}$,
where $M^{\prime }$ denotes a scale of physics beyond the Grand Unification
scale. Taking this to be the Planck scale (probably the largest reasonable
possibility) and $<H^{I_{W,R}=1/2}>$ to be the Grand Unified scale (it
breaks any Grand Unified group) one finds 
\[
M_{M} ~\simeq~ \frac{(10^{16})^{2}}{10^{19}} ~{\rm GeV } 
\]
giving 
\[
m_{Light} ~\simeq~ 1 ~{\rm eV } 
\]
Thus, one sees that the see saw mechanism naturally gives neutrino masses in
the range relevant to neutrino oscillation measurements. Moreover, as the
neutrino mass is proportional to the Dirac mass squared, taking the Dirac
mass of each family of neutrinos to be of the order of the equivalent up
quark mass, one obtains a large hierarchy between different families of
neutrino. This is what is required if one is to explain several oscillation
phenomena, for it allows the existence of several mass differences.

In what follows, we will concentrate on the possibility that there is a
minimal extension of the Standard Model involving just three new
right-handed neutrino states and that the mass structure of the neutrinos is
intimately related to that of the charged leptons and quarks. This implies
that the three different indications for neutrino oscillations discussed
above cannot be simultaneously explained, because three neutrino masses
allow only two independent mass differences. To explain all three
observations requires another (sterile) light neutrino state. However,
introducing such a state breaks any simple connection between neutrino
masses and those of the other Standard Model states and here we wish to
explore whether the apparently complex pattern of quark and lepton masses
and mixing angles can \ be simply understood. In this, while the structure
of the see-saw mechanism leads naturally to light neutrino masses in a
physically interesting range, it does not by itself explain the pattern of
neutrino masses and mixing angles.

To go further requires some family symmetry capable of relating the masses
of different generations. The recent Super-Kamiokande measurements have
triggered a large amount of work studying the implications for neutrino
masses in extensions of the fundamental theory \cite{recent}. Actually, the
origin of fermion masses and mixing angles, including those of neutrinos,
has been studied in numerous publications \cite{FN,textures,IR}. An obvious
possibility is that the various hierarchies arise due to some symmetry at a
higher scale. An indication that additional symmetries exist, has been
provided by the observation that the fermion mixing angles and masses have
values consistent with the appearance of ``texture'' zeros in the mass
matrices. In this framework, the predictions for neutrino textures in models
have been studied in \cite{neutr,DLLRS,LLR,ver}. In many cases, a large
mixing angle is not easy to reproduce, principally because of the
constrained form of the Dirac mass matrices \cite{DLLRS}. However, in
certain cases, the Dirac sector may lead naturally to such a large mixing,
as we showed in~\cite{LLR} (similar conclusions were recently discussed in 
\cite{BPW}). Here, we revisit these models in the light of the recent
results and study the expected predictions in more detail. In order to avoid
the hierarchy problem that is associated with the large mass scale necessary
for the see-saw mechanism, we assume the Standard Model descends from a
supersymmetric theory with a low scale of supersymmetry breaking.

\section{Family symmetry and hierarchical quark and lepton masses}

It has been observed \cite{FN,textures} that the hierarchical structure for
the fermion mass matrices strongly suggests it originates from a
spontaneously broken family symmetry. In this approach, when the family
symmetry is exact, only the third generation will be massive corresponding
to only the (3,3) entry of the mass matrix being non-zero. When the symmetry
is spontaneously broken, the zero elements are filled in at a level
determined by the symmetry. Suppose a field \ $\theta $ which transforms
non-trivially under the family symmetry acquires a vacuum expectation value,
thus spontaneously breaking the family symmetry. The zero elements in the
mass matrix will now become non-zero at some order in $<\theta >.$ If only
the 2-3 and 3-2 elements are allowed by the symmetry at order $\theta /M,$
where $M$ is a mass scale to be determined, then a second fermion mass will
be generated at $O((\theta /M)^{2}).$ In this way one may build up an
hierarchy of masses. 
\begin{equation}
{\cal M}\sim \left( 
\begin{array}{ccc}
0 & 0 & 0 \\ 
0 & 0 & 0 \\ 
0 & 0 & 1
\end{array}
\right) \rightarrow \left( 
\begin{array}{ccc}
0 & 0 & 0 \\ 
0 & 0 & <\theta >/M \\ 
0 & <\theta >/M & 1
\end{array}
\right)
\end{equation}
How do these elements at $O(\theta /M)$ arise? A widely studied approach
communicates symmetry breaking via an extension of the ``see-saw'' mechanism
mixing light to heavy states - in this context it is known as the Froggatt
Nielsen mechanism \cite{FN}. To illustrate the mechanism, suppose there is a
vector-like pair of quark states $X$ and $\overline{X}$ with mass $M$ and
carrying the same Standard Model quantum numbers as the $c_{R}$ quark, but
transforming differently under the family symmetry, so that the Yukawa
coupling $h\overline{c_{L}}XH$ is allowed. Here $H$ is the Standard Model
Higgs responsible for giving up quarks a mass. When $H$ acquires a vacuum
expectation value (vev), there will be mixing between $\overline{c_{L}}$ and 
$\overline{X}.$ If in addition there is a gauge singlet field $\theta $
transforming non-trivially under the family symmetry so that the coupling $%
h^{\prime }\overline{X}c_{R}\theta $ is allowed, then the mixing with heavy
states will generate the mass matrix.

\[
\left( 
\begin{array}{cc}
\overline{c_{L}} & \overline{X}
\end{array}
\right) \left( 
\begin{array}{cc}
0 & h<H> \\ 
h^{\prime }<\theta > & M
\end{array}
\right) \left( 
\begin{array}{c}
c_{R} \\ 
X
\end{array}
\right) 
\]
Diagonalising this gives a see-saw mass formula 
\begin{equation}
m_{c}\simeq \frac{hh^{\prime }<H><\theta >}{M}  \label{mc}
\end{equation}
This mass arises through mixing of the light with heavy quarks.

A similar mechanism can generate the mass through mixing of the light Higgs
with heavy Higgs states. Suppose $H_{X},$ $\overline{H}_{X}$ are Higgs
doublets with mass $M.$ If $H_{X}$ has family quantum numbers allowing the
coupling $H\overline{H}_{X}\theta $, there will be mixing between $H$ and $%
H_{X}.$ If the family symmetry also allows the coupling $\overline{c_{L}}%
c_{R}H_{X}$, the light-heavy Higgs mixing induces a mass for the charm quark
of the form given in eq.(\ref{mc}).

\begin{figure}[h]
\centerline{\epsfig{figure=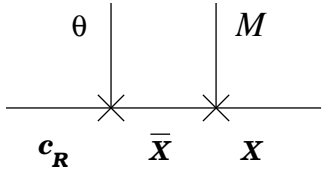,width=0.25\textwidth,clip=}}
\caption{Generation of non-renormalisable operators through quark mixing}
\end{figure}

\vspace*{0.2 cm}

\begin{figure}[h]
\centerline{\epsfig{figure=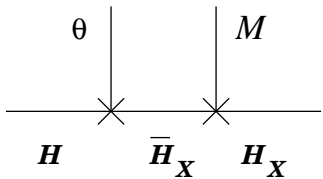,width=0.25\textwidth,clip=}}
\caption{Generation of non-renormalisable operators through Higgs mixing}
\end{figure}

\vspace*{0.1 cm}

\subsection{Abelian family symmetry\label{sec:abelian}}

\bigskip 
\begin{table}[tbp] \centering%
%
\begin{tabular}{|l|}
\hline
{\small 
\begin{tabular}{|c|cccccccc|}
\hline
& $Q_{i}$ & $u_{i}^{c}$ & $d_{i}^{c}$ & $L_{i}$ & $e_{i}^{c}$ & $\nu
_{i}^{c} $ & $H_{2}$ & $H_{1}$ \\ \hline
$U(1)_{FD}$ & $\alpha _{i}$ & $\beta _{i}$ & $\gamma _{i}$ & $b_{i}$ & $%
c_{i} $ & $d_{i}$ & $-\alpha _{3}-\beta _{3}$ & $-\alpha _{3}-\gamma _{3}$
\\ \hline
\end{tabular}
} \\ \hline
\end{tabular}
\caption{$U(1)_{FD}$ charges\label{table:1}}%
\end{table}%
%

How difficult is it to find a family symmetry capable of generating an
acceptable fermion mass matrix? The surprising answer is ``Not at all
difficult'' and the simplest possibility using an Abelian family symmetry
group \cite{IR,horiz1} works very well.

As a bonus, such symmetries can also give texture zeros simultaneously in
the (1,1) and (1,3) positions, generating good predictions relating masses
and mixing angles \cite{IR}. The basic idea is that the structure of the
mass matrices is determined by a flavour dependent family symmetry, $%
U(1)_{FD}$. The most general charge assignments of the various states under
this symmetry are given in Table \ref{table:1}. If the light Higgs, $H_{2}$, 
$H_{1}$, that generate the up-quark (Dirac neutrino) and down-quark (charged
lepton) masses respectively, have $U(1)$ charge so that only the (3,3)
renormalisable Yukawa coupling to $H_{2}$, $H_{1}$ is allowed, then only the
(3,3) element of the associated mass matrix will be non-zero. The remaining
entries are generated when the $U(1)$ symmetry is broken. This breaking is
taken to be spontaneous via \sm singlet fields carrying family charge
acquiring vacuum expectation values (vevs). For example the fields, $\theta
,\;\bar{\theta}$, with $U(1)_{FD}$ charge -1, +1 respectively \footnote{%
In some models only fields with one sign of family charge acquire vevs. In
this case holomorphic zeros may occur in the mass matrices as is discussed
below.}, may acquire equal vevs along $D-$flat directions. After this
breaking, the structure of the mass matrices is generated. Let us discuss,
as an example, the origin of the (3,2) entry in the up quark mass matrix.
This appears at order $\epsilon ^{\mid \alpha _{2}-\alpha _{3}\mid }$
because U(1) charge conservation allows only the non-renormalisable operator 
$c^{c}tH_{2}(\theta /M_{2})^{\beta _{2}-\alpha _{3}},\;\beta _{2}>\alpha
_{3} $ or $c^{c}tH_{2}(\bar{\theta}/M_{2})^{\alpha _{3}-\beta _{2}},\;\alpha
_{3}>\beta _{2}$. Here $\epsilon =(<\theta >/M_{2})$ and $M_{2}$ is the
unification mass scale which governs the higher dimension operators. As we
discussed above this is most likely to be the mass of the heavy quark or
heavy Higgs which, on spontaneous breaking of the family symmetry, mixes
with the light states.

\subsection{Abelian family symmetry and large lepton mixing\label{sec:2x2}}

Let us consider in more detail the $2\times 2$ heavier sector of the theory,
relevant to the atmospheric neutrino oscillations for the case only one mass
squared difference contributes. The charged lepton matrix constrained by the
U(1) family symmetry has the form 
\begin{equation}
\frac{{\cal M}_{\ell }}{m_{\tau }}=\left( 
\begin{array}{cc}
c\left( \frac{\theta }{M}\right) ^{q_{L}}\left( \frac{\theta }{M}\right)
^{q_{R}} & a\left( \frac{\theta }{M}\right) ^{q_{L}} \\ 
b\left( \frac{\theta }{M}\right) ^{q_{R}} & 1
\end{array}
\right)  \label{mform}
\end{equation}
where the origin of the intermediate mass scale, $M,$ will be discussed
shortly. The parameters $a,~b,~c$ are constants of O(1), reflecting the
unknown Yukawa couplings and $q_{L}=b_{2}-b_{3}$, $q_{R}=c_{2}-c_{3}%
\footnote{%
Here, for simplicity, we assume $b_{i},$ $c_{i},$ $q_{L},$ $q_{R}$ are all
positive. The analysis also applies without this restriction for the case $%
<\theta >\approx <\overline{\theta }>.$ We will discuss what happens when
these conditions are not satisfied later.}$.

It is instructive to write this in the form 
\begin{equation}
\frac{{\cal M}_{\ell }}{m_{\tau }}=E\left( (\frac{\theta }{M}%
)^{q_{L}}\right) \cdot A\cdot E\left( (\frac{\theta }{M})^{q_{R}}\right)
\label{mform1}
\end{equation}
where 
\begin{equation}
E(x)=\left( 
\begin{array}{cc}
x & 0 \\ 
0 & 1
\end{array}
\right) ,~~A=\left( 
\begin{array}{cc}
c & a \\ 
b & 1
\end{array}
\right) ,  \label{Edef}
\end{equation}
The matrix $A$ is determined by the Yukawa couplings only. If the only
symmetry restricting the form of the mass matrices is the Abelian family
symmetry there is no reason to expect correlations between the elements of $%
A $ and so we expect $Det(A)=O(1).$ This is the situation we will explore in
this paper. Given this we may see that ${\cal M}_{\ell }^{D}$ has the form 
\begin{equation}
{\cal M}_{\ell }=V_{\ell L}\cdot {\cal M}_{\ell ,Diagonal}\cdot V_{\ell
R}^{T}
\end{equation}
where 
\begin{equation}
\frac{{\cal M}_{\ell ,Diagonal}}{m_{\tau }}=\left( 
\begin{array}{cc}
r(\frac{\theta }{M})^{q_{L}}(\frac{\theta }{M})^{q_{R}} & 0 \\ 
0 & 1
\end{array}
\right)
\end{equation}
and 
\begin{equation}
V_{\ell L}=V\left( r^{\prime }(\frac{\theta }{M})^{q_{L}}\right) ,~~V_{\ell
R}=V\left( r^{\prime \prime }(\frac{\theta }{M})^{q_{R}}\right)
\end{equation}
with $r,~r^{\prime },r^{\prime \prime }=O(1)$ and $V(x)=\left( 
\begin{array}{cc}
1 & x \\ 
-x & 1
\end{array}
\right) $. The lepton analogue \cite{mns} of the CKM mixing matrix for
quarks is given by 
\begin{equation}
V_{MNS}\approx V_{\nu L}^{\dagger }\cdot V_{\ell L}  \label{MM}
\end{equation}

The important point to note is that the left-handed lepton mixing matrix
contribution is determined entirely by the {\it left-handed} lepton doublet
family symmetry charges while the eigenvalues are determined by both the
left-handed and right-handed charges.

A similar analysis may be applied to the neutrino sector. We have 
\begin{eqnarray}
{\cal M}_{\nu }^{effective} &=&{\cal M}_{\nu }^{D}\cdot ({\cal M}_{\nu
}^{M})^{-1}\cdot {\cal M}_{\nu }^{DT} \\
&=&(V_{\nu L}\cdot {\cal M}_{\nu ,Diagonal}^{D}\cdot V_{\nu R}^{(D)T})\cdot (%
{\cal M}_{\nu }^{M})^{-1}\cdot (V_{\nu L}\cdot {\cal M}_{\nu
,Diagonal}^{D}\cdot V_{\nu R}^{(D)T})^{T}  \nonumber \\
&\equiv &V_{\nu L}\cdot V_{\nu R}\cdot {\cal M}_{\nu
,Diagonal}^{effective}\cdot V_{\nu R}^{T}\cdot V_{\nu L}^{T}  \nonumber
\end{eqnarray}
We see that there are two contributions to $V_{MNS}$ in eq.(\ref{MM}) coming
from the neutrino sector. The first is 
\begin{equation}
V_{\nu L}=V\left( s(\frac{\theta }{M^{\prime }})^{q_{L}}\right)
\end{equation}
where $s=O(1)$ and we have allowed for a different intermediate scale $%
M^{\prime }$ (see below). It is determined by the same {\it left-handed}
lepton doublet family symmetry charges that determine $V_{\ell L}.$

The second contribution, $V_{\nu R},$ is sensitive to the right-handed
neutrino family charges. However in the case the light neutrinos have a
hierarchical mass pattern (necessary if we are to explain both the
atmospheric and solar oscillations) this contribution cannot be large. To
see this note that if the elements of $V_{\nu R}$ are all of $O(1)$ and one
neutrino mass, $m_{1},$ dominates then the elements of the matrix $V_{\nu
R}\cdot {\cal M}_{\nu ,Diagonal}^{effective}\cdot V_{\nu R}^{T}$ are all of $%
O(m_{1})$ but its determinant is $<<O(m_{1}^{2}{)}.$ This matrix is also
given by (${\cal M}_{\nu ,Diagonal}^{D}\cdot V_{\nu R}^{(D)T})\cdot ({\cal M}%
_{\nu }^{M})^{-1}\cdot ({\cal M}_{\nu ,Diagonal}^{D}\cdot V_{\nu
R}^{(D)T})^{T}.$ As discussed above, the Abelian family symmetry cannot give
correlations between the Yukawa couplings determining different matrix
elements of ${\cal M}_{\nu ,Diagonal}^{D}$ and ${\cal M}_{\nu }^{M}.$ Thus,
its determinant cannot be of a different order than the product of its
diagonal elements, in contradiction with the conclusion that follows if the
neutrinos are hierarchical in mass. The implication is that large mixing can
only come from the right-handed neutrino sector if there are two nearly
degenerate neutrinos. If we are to describe solar neutrino mixing too, this
has to be extended to three nearly degenerate neutrinos \cite{bhrk} and
since an Abelian symmetry alone cannot generate this structure we dismiss
this possibility here. As a result, we require a hierarchical neutrino mass
pattern and this implies $V_{MNS}\approx V_{\nu L}^{T}\cdot V_{\ell L}$
giving 
\begin{equation}
\sin \theta _{\mu \nu _{\tau }}\approx r^{\prime }(\frac{\theta }{M}%
)^{q_{L}}-s(\frac{\theta }{M^{\prime }})^{q_{L}}  \label{MM1}
\end{equation}
with the implication that $q_{L}=0$ for near maximal mixing.

At this point, it is important to discuss what are the expansion parameters
in the various sectors, i.e. what are $M,$ and $M^{\prime }$ . As discussed
above, the most reasonable origin of the higher dimension terms $\propto
\left( \frac{\theta }{M_{i}}\right) ^{a}$ is via the Froggatt-Nielsen
mechanism \cite{FN}, through the mixing of the lepton states or the Higgs
states. In the case of the mixing responsible for $V_{MNS}$, the former is
irrelevant for in this case the mixing arises via heavy states which belong
to $SU(2)$ doublets and hence are closely degenerate $(M^{\prime }=M).$ In
this case, the contributions to eq.(\ref{MM}) or (\ref{MM1}) cancel. We
conclude that {\em the relevant mixing is generated through the Higgs states}%
. Thus, $M$ should be interpreted as the mass of the heavy Higgs states
mixing with $H_{1}$, generating the down quark and charged lepton masses,
while $M^{\prime }$ is the mass of the Heavy Higgs state mixing with $H_{2}$%
, generating the up quark masses. Consequently, the expectation is that $%
M_{1}>M_{2}$ because the same expansion parameters govern the hierarchy of
quark masses, and typically one needs a smaller expansion parameter in the
up-quark sector to explain the larger hierarchy of masses in that sector.
This in turn implies that {\em the lepton mixing comes primarily from the
charged lepton sector}. Of course, this conclusion depends on the relative
up and down-quark charges - we will return in a discussion of this shortly.

Although we have argued that the mixing matrix $V_{MNS}$ is determined by
the left-handed charges only, the mass eigenvalues are sensitive to the
right-handed charges. In particular the Majorana mass has a similar form to
that in eq.(\ref{mform1}) 
\begin{equation}
{\cal M}_{\nu }^{M}\varpropto E\left( (\frac{\theta }{M^{\prime \prime }}%
)^{q_{R}}\right) \cdot B\cdot E\left( (\frac{\theta }{M^{\prime \prime }}%
)^{q_{R}}\right)
\end{equation}
where we have allowed for a different intermediate mass scale, $M^{\prime
\prime },$ in the right-handed sector and $B$ is a matrix of Yukawa
couplings of $O(1).$ This gives 
\begin{equation}
Det(m_{eff})=\frac{[Det(M_{\nu }^{D})]^{2}}{Det(M_{\nu }^{R})}\varpropto 
\frac{(\frac{\theta }{M^{\prime }})^{2q_{R}}(\frac{\theta }{M})^{2q_{L}}}{(%
\frac{\theta }{M^{\prime \prime }})^{2q_{R}}}
\end{equation}
To summarise, the choice $q_{L}=0$ leads to $O(1)$ mixing, although there is
no reason for the mixing to be really maximal i.e. $\pi /4$ (for this, a
non-Abelian symmetry is necessary \cite{bhrk}). The lepton mass may be
adjusted by the choice of $q_{\ell R}$, while the neutrino masses may be
adjusted by the choice of $q_{\nu R}$. Thus, a $U(1)$ family symmetry is
readily compatible with an hierarchical neutrino mass matrix and a large
mixing angle in the lepton sector although it is unlikely to be maximal.

At this point it is perhaps useful to comment how {\it maximal} mixing {\it %
can} be obtained from the right-handed neutrino sector \cite
{ver,allanach,shafi} via an Abelian symmetry. This may be arranged using the
holomorphic structure of the superpotential in supersymmetric theories.
Suppose the $<\overline{\theta }>=0$ i.e. only the positive family charge
field $\theta $ acquires a vev. Suppose further the family charges of the
heavy lepton doublets are given by $q_{\mu _{L}}=n,$ $q_{\tau _{L}}=0$ while
the right-handed neutrinos have charge given by $q_{\nu _{\mu }}=(p+r)$ $%
q_{\nu _{\tau }}=-p$ with $n,$ $p$ and $r$ positive and $n\eqslantgtr p$.\
Then we have 
\begin{eqnarray}
{\cal M}_{\nu }^{D} &\varpropto &\left( 
\begin{array}{cc}
(\frac{\theta }{M})^{n+p+r} & (\frac{\theta }{M})^{n-p} \\ 
(\frac{\theta }{M})^{p+r} & 0
\end{array}
\right) \\
{\cal M}_{\nu }^{M} &\varpropto &\left( 
\begin{array}{cc}
(\frac{\theta }{M^{\prime }})^{2p+r} & 1 \\ 
1 & 0
\end{array}
\right)  \nonumber
\end{eqnarray}
The zeros here arise because the net charge in the $(2,2)$ element is
negative (the case not allowed in the discussion above). For the simple case 
$M=M^{\prime }$ we have 
\begin{eqnarray}
{\cal M}_{\nu }^{effective} &=&{\cal M}_{\nu }^{D}\cdot ({\cal M}_{\nu
}^{M})^{-1}\cdot {\cal M}_{\nu }^{DT} \\
&\varpropto &\left( 
\begin{array}{cc}
(\frac{\theta }{M})^{n} & 1 \\ 
1 & 0
\end{array}
\right)  \nonumber
\end{eqnarray}
(a similar structure applies for a range of $M/M^{\prime })$. This gives
maximal mixing and two nearly degenerate neutrinos. Thus if we wish to
describe solar as well as atmospheric oscillations it is necessary to add a
sterile neutrino \cite{shafi,ster}. We do not consider such schemes
here.\bigskip

\section{Gauge unification constraints}

While an Abelian family symmetry provides a promising origin for an
hierarchical pattern of fermion masses, in order to go further it is
necessary to specify the charges of the quarks, charged leptons and
neutrinos. As we discussed in the last section, it is straightforward to fit
all the observed masses and mixing angles by the choice of the $U(1)$
charges not constrained by the Standard Model gauge symmetry. However, the
structure of the Standard Model is suggestive of an underlying unification
which may relate quark and lepton multiplets. The success of the unification
of the gauge couplings also supports this picture. Thus, we think it of
interest to consider whether realistic quark mass structures are consistent
with the constraints on an Abelian family symmetry that result from some
underlying unified gauge symmetry.

\subsection{$SO(10)\times U(1)\label{so10s}$}

Consider first the possibility that the family symmetry commutes with an $%
SO(10)$ GUT. In this case, all quark and lepton charges for the left- and
charge conjugate right-handed fields in a given generation are the same.
This gives rise to a left-right symmetric mass matrix with similar structure
for the up quark, the down quark, the charged lepton and the Dirac neutrino
mass matrices. The only difference between these sectors is due to the
possibility the expansion parameters can be different. Thus $SO(10)$
provides a specific realisation of the first model of lepton masses that has
been discussed in reference \cite{IR} for the case $b=0$. Following the
discussion of Section \ref{sec:2x2} we note that the expansion parameters
determining $V_{MKS}$ are principally those arising from Higgs mixing. Since
the same Higgs is responsible for the structure of the down quark mass and
the charged lepton mass this leads to the prediction 
\begin{equation}
V_{\mu {\tau }}\approx V_{cb}  \label{so10}
\end{equation}
Clearly this is in gross conflict with observations so to rescue it it is
necessary for the coefficients of $O(1)$ associated with the down quark and
leton sectors to differ. In our analysis, we are going to discard solutions
that, in order to match the observations, require the existence of either
cancellations that are not predicted by the Abelian family symmetry, or
coefficients with magnitude comparable to that of the expansion parameter
ordering the elements of the mass matrices \footnote{%
This requirement is unreasonable in the case the down quark and lepton
couplings are {\it predicted} to differ by the underlying GUT. We consider
such a possibility for the case of $SU(5)$ in the next Section but choose
not to pursue it for $SO(10).$}. For this reason we do not consider this $%
SO(10)$ possibility further. We also apply these criteria to the analysis in
the rest of this section.

We should further stress that in our analysis we use the GUT structure only
in order to constrain the $U(1)$ flavour charges of the light fields. In
particular we assume that all terms allowed at a given order by the family
symmetry do in fact occur. This condition can be avoided if the heavy fields
responsible for the Froggatt Nielsen mixing have restricted $U(1)$ family
charges. A simple example of this mechanism appears in \cite{IR};  viable $%
SO(10)$ examples appear in \cite{SO10a,SO10b}.  We do not consider such a
possibility here, because we wish to explore whether the $U(1)$
family symmetry structure of the light fields alone is sufficient to
determine the pattern of light fermion masses and mixings.

\subsection{$SU(5)\times U(1)$}

We turn to the possibility that the family symmetry commutes with an $SU(5)$
GUT. This is, of course, consistent with an underlying $SO(10)$ structure
but to avoid the bad relation of eq.(\ref{so10}) it is necessary for the
Abelian family symmetry to have a component along the $SO(10)$ neutral
generator $\varpropto (B-L)$ which commutes with $SU(5).$ There are only
three $U(1)$ family charges needed for each family. These are given by 
\begin{eqnarray}
Q_{(q,u^{c},e^{c})_{i}} &=&Q_{i}^{10}  \label{su5charges} \\
Q_{(l,d^{c})_{i}} &=&Q_{i}^{\overline{5}}  \nonumber \\
Q_{(\nu _{R})_{i}} &=&Q_{i}^{\nu _{R}}  \nonumber
\end{eqnarray}

From the above it immediately follows that :

(i) The up-quark mass matrix is symmetric.

(ii) the charged lepton mass matrix is the transpose of the down quark mass
matrix.

The expansion parameters in the various sectors can be different (depending
on whether the non-renormalisable contributions are due to fermion or to
Higgs mixing, or a to combination of the two). However, as discussed in
Section \ref{sec:2x2}, a single expansion parameter describing $H_{1}$
mixing determines the down quark and charged lepton {\it mixing}, and
similarly a single expansion parameter describing $H_{2}$ mixing determines
the up quark and Dirac neutrino {\it mixing}. The fact that the right-handed
neutrino charges are unconstrained means the neutrino mass spectrum is not
restricted but, again as discussed in Section \ref{sec:2x2}, the mixing
angle in the $\mu \tau $ sector is insensitive to these charges and is
determined primarily by $Q_{\mu ,\tau }^{\overline{5}}.$

At first sight this charge structure seems to offer an immediate explanation
for the difference between large mixing angle observed in atmospheric
neutrino mixing and the small quark mixing angles. This is because the
former is determined by $Q_{\mu ,\tau }^{\overline{5}}$ while the
corresponding quark mixing matrix element, $V_{cb},$ is determined by $%
Q_{i}^{10}.$ However the main difficulty in using this freedom to describe
both mixings arises from the associated correlations between the eigenvalues
of the charged lepton and the down quark mass matrices due to structure (ii)
above. Indeed, if the eigenvalues of the down mass matrix (with expansion
parameter $e)$ are given by a sequence $1,e,e^{k}$ the eigenvalues for the
leptons (with expansion parameter $\tilde{\epsilon})$ are $1,\tilde{\epsilon}%
,\tilde{\epsilon}^{k}$. The down quark masses are well described by the
choice $e\simeq \overline{\epsilon }^{2}\simeq 0.04$ and $k=2.$ while for
the leptons the hierarchies are well described by $\tilde{\epsilon}\simeq 
\bar{\epsilon}\simeq 0.2$ and $k=5.$ This is clearly inconsistent with the
pattern coming from the family symmetry which requires the same $k$ in the
down quark and lepton sectors.

One way to reconcile the two forms for the mass matrix, originally advocated
by Georgi and Jarlskog \cite{gj}, is to have different Yukawa couplings in
the quark and lepton sectors. These couplings are determined by the
underlying $SU(5)$ gauge group. If the mass comes from the coupling to a $5$
of Higgs then $m_{d_{i}}=m_{l_{i}}$ while if the mass comes from the
coupling to a $45$ of Higgs then $m_{d_{i}}=3m_{l_{i}}.$The observed
hierarchy for the lepton masses is well described by the eigenvalues $1,3%
\bar{\epsilon}^{2},\bar{\epsilon}^{4}/3$. Georgi and Jarlskog achieved this
by restricting the mass matrices by family symmetries to have the form 
\begin{eqnarray}
{\cal M}_{d} &=&\left( 
\begin{array}{ccc}
0 & a^{\prime }<H^{5}> & 0 \\ 
a<H^{5}> & c<H^{45}> & 0 \\ 
0 & 0 & b<H^{5}>
\end{array}
\right)  \label{gj} \\
{\cal M}_{\ell } &=&\left( 
\begin{array}{ccc}
0 & a^{\prime }<H^{5}> & 0 \\ 
a<H^{5}> & 3c<H^{45}> & 0 \\ 
0 & 0 & b<H^{5}>
\end{array}
\right)  \nonumber
\end{eqnarray}
While this gives an acceptable pattern of masses it clearly does not give
the large mixing angle in the $\mu \tau $ sector. Here we will determine
whether it is possible in more general schemes. To do so, we look explicitly
at the forms of the matrices. Given that the top quark is very heavy it is
reasonable to assume that it is given by an $O(1)$ renormalisable
contribution. Then, the up-quark mass matrix is specified to be: 
\begin{equation}
{\cal M}_{u}\propto \left( 
\begin{array}{ccc}
{\epsilon }^{|2x|} & {\epsilon }^{|x+b|} & {\epsilon }^{|x|} \\ 
{\epsilon }^{|x+b|} & \epsilon ^{|2b|} & \epsilon ^{|b|} \\ 
\epsilon ^{|x|} & \epsilon ^{|b|} & 1
\end{array}
\right)  \label{tsu5}
\end{equation}
where $x=Q_{1}^{10}-Q_{3}^{10}$ and $b=Q_{2}^{10}-Q_{3}^{10}$. Then $\frac{%
m_{c}}{m_{t}}=e^{|2b|}$ and $e^{|b|}\approx 0.045$ gives a good fit. The
contribution of the up sector to $V_{cb}$ is given by $%
V_{cb}^{up}=e^{|b|}=0.045.$

What about the light up-quark hierarchies? We have

\begin{equation}
\frac{m_{u}}{m_{c}}=max(\frac{e^{|x+b|}}{e^{|2b|}},\frac{e^{|x|}}{e^{|b|}}),
\end{equation}
indicating that either 
\begin{equation}
e^{\;|x+b|-2|b|\;}={\cal O}(10^{-6})
\end{equation}
or 
\begin{equation}
e^{|x|-|b|\;}={\cal O}(10^{-6})
\end{equation}

We now pass to the down-quark and charged lepton hierarchies. We consider
the case that the (3,3) element of the quark and lepton mass matrices is
allowed by the family symmetry and the difference between the top and the
bottom quark masses is largely due to $tan\beta \equiv <H_{1}>/<H_{2}>$
being large. In this case the charge of $H_{2}$ is fixed to be the same as
of $H_{1}$. Then, the down quark and charged lepton textures have the form :

\begin{equation}
M_{\ell }\propto {\tilde{\epsilon}}^{|Q_{3}^{\overline{5}}-Q_{3}^{10}|}%
\left( 
\begin{array}{ccc}
{\tilde{\epsilon}}^{|x+y|} & {\tilde{\epsilon}}^{|y+b|} & {\tilde{\epsilon}}%
^{|y|} \\ 
{\tilde{\epsilon}}^{|x+a|} & \tilde{\epsilon}^{|a+b|} & \tilde{\epsilon}%
^{|a|} \\ 
\tilde{\epsilon}^{|x|} & \tilde{\epsilon}^{|b|} & 1
\end{array}
\right) ,\;\;\;M_{down}\propto \bar{\epsilon}^{|Q_{3}^{\overline{5}%
}-Q_{3}^{10}}\left( 
\begin{array}{ccc}
{\bar{\epsilon}}^{|x+y|} & {\bar{\epsilon}}^{|x+a|} & {\bar{\epsilon}}^{|x|}
\\ 
{\bar{\epsilon}}^{|y+b|} & \bar{\epsilon}^{|a+b|} & \bar{\epsilon}^{|b|} \\ 
\bar{\epsilon}^{|y|} & \bar{\epsilon}^{|a|} & 1
\end{array}
\right)
\end{equation}
where $a=Q_{2}^{\overline{5}}-Q_{3}^{\overline{5}}$ and $y=Q_{2}^{\overline{5%
}}-Q_{3}^{\overline{5}}.$

From the above matrices, we see that the eigenvalues for the lepton mass
matrix are 
\begin{equation}
1,\tilde{\epsilon}^{|a+b|},max(\tilde{\epsilon}^{|x+y|},\frac{\tilde{\epsilon%
}^{|x+a|}\tilde{\epsilon}^{|y+b|}}{\tilde{\epsilon}^{|a|}\tilde{\epsilon}%
^{|b|}}),
\end{equation}
while for the down quarks 
\begin{equation}
1,\bar{\epsilon}^{|a+b|},max(\bar{\epsilon}^{|x+y|},\frac{\bar{\epsilon}%
^{|x+a|}\bar{\epsilon}^{|y+b|}}{\bar{\epsilon}^{|a|}\bar{\epsilon}^{|b|}})
\end{equation}
Clearly, irrespective of the choice of expansion parameters, this form will
not simultaneously generate the correct ratios for down quarks and leptons
without requiring the lepton Yukawa couplings are different from the quark
Yukawa couplings. Following the suggestion of Georgi and Jarlskog we assume
the Higgs responsible for the (2,2) entry is a $45$ representation of $%
SU(5). $ This generates a relative factor of 3 in the (22) entry of the
lepton mass matrix. Then provided that the (11) entry is smaller than the
(12) entry, the smallest eigenvalue is suppressed by a factor of 3. For this
to occur, we need: $xy>0$, while $|y+b|+|x+a|-|a|-|b|$ has to be smaller
than $x+y$.

Can we reconcile these constraints with an acceptable pattern of mixing
angles? Let us first consider the most predictive case which has texture
zeros in the $(1,1)$ and $(1,3)$ positions (see Section \ref{sec:tz} and
ref. \cite{IR}). In this case the left-handed quarks have charges $-4,1,0$
and $V_{CKM}^{12}$ has to arise from the down sector. If we want
near-maximal lepton mixing, the two heavier {\it right}-handed down charges
(which are the same as those of the two heavier {\it left}-handed lepton
charges respectively) have to be zero. In this case, the second generation
left-handed quark charge, $|\alpha _{2}|=1$, give an unacceptably large $%
m_{s}/m_{b}$.

To obtain the correct $m_{s}/m_{b}$ for the case of maximal lepton mixing we
must give up the texture zero structure and choose $|\alpha _{2}|=2$ and $%
|\alpha _{3}|=0$. Then, to obtain the correct (12)-quark mixing from the
down sector (note that to get such a large mixing from the up sector is more
difficult, as it would lead to a large up-quark mass), we need $|\alpha
_{1}|=3$. What does this imply for $M_{u}$? For $\alpha _{1}=3$, the mass of
the up-quark is given by 
\begin{equation}
\frac{m_{up}}{m_{c}}\approx \frac{M_{u}(1,1)-M_{u}^{2}(1,2)/M_{u}(2,2)}{%
M_{u}(2,2)}\approx O(\epsilon ^{6}-\epsilon ^{6})  \label{canc}
\end{equation}
where now $\epsilon =0.23$ (this is in order to obtain the correct charm
mass, for our choice $|\alpha _{2}|=2$). Thus we see that there must be a
cancellation between the two terms, or the introduction of small
coefficients to obtain the correct ratio for these masses.

Let us now go back to the down mass matrix. The only charge undetermined is
that for $d_{1}^{c}$. Fixing this to +1, gives the correct mass for the down
quark.

The obtained mass matrices are: 
\begin{equation}
\frac{M_{u}}{m_{t}}=\left( 
\begin{array}{ccc}
\bar{\epsilon}^{6} & \bar{\epsilon}^{5} & \bar{\epsilon}^{3} \\ 
\bar{\epsilon}^{5} & \bar{\epsilon}^{4} & \bar{\epsilon}^{2} \\ 
\bar{\epsilon}^{3} & \bar{\epsilon}^{2} & 1
\end{array}
\right) ,\frac{M_{down}}{m_{b}}=\left( 
\begin{array}{ccc}
\bar{\epsilon}^{4} & \bar{\epsilon}^{3} & \bar{\epsilon}^{3} \\ 
\bar{\epsilon}^{3} & \bar{\epsilon}^{2} & \bar{\epsilon}^{2} \\ 
\bar{\epsilon} & 1 & 1
\end{array}
\right) ,\frac{M_{\ell }}{m_{\tau }}=\left( 
\begin{array}{ccc}
\bar{\epsilon}^{4} & \bar{\epsilon}^{3} & \bar{\epsilon} \\ 
\bar{\epsilon}^{3} & \bar{\epsilon}^{2} & 1 \\ 
\bar{\epsilon}^{3} & \bar{\epsilon}^{2} & 1
\end{array}
\right) ,
\end{equation}

This is the choice of charges that appears in \cite{af} (with the exception
of the down and lepton charges of the first generation). To summarise, we
see that within the framework of $SU(5)$, viable solutions only exist
provided (c.f. eq.(\ref{canc})~) the $O(1)$ coefficients have a special form
not guaranteed by the Abelian symmetry applied to the light fields alone
(c.f the discussion in Section 3.1). Since here we are concerned to explore
how much of the fermion mass patterns can be generated by the Abelian
symmetry alone, we will not discuss these $SU(5)\times U(1)$ models further.

\subsection{Flipped $SU(5)\times U(1)$}

In the case of the flipped $SU(5)$, the fields $Q_{i},d_{i}^{c}$ and $\nu
_{i}^{c}$ belong to $10${\bf \ }of{\bf \ }$SU(5)$, while $u_{i}^{c}$ and $%
L_{i}$ belong to a $\overline{5}$. Finally the $e_{i}^{c}$ fields belong to
singlet representations of $SU(5)$.

The above assignment, implies that the down quark mass matrices are
symmetric, and therefore they are expected to have the form presented in 
\cite{IR}. Then we obtain viable hierarchies by fixing the down-quark
charges to ie $4,-1,0$ and the expansion parameter in the down mass matrix
to be $\bar{\epsilon}=0.23$. 
\begin{equation}
M_{down}\propto \left( 
\begin{array}{ccc}
\bar{\epsilon}^{8} & \bar{\epsilon}^{3} & \bar{\epsilon}^{4} \\ 
\bar{\epsilon}^{3} & \bar{\epsilon}^{2} & \bar{\epsilon} \\ 
\bar{\epsilon}^{4} & \bar{\epsilon} & 1
\end{array}
\right)  \label{mm}
\end{equation}
Since the charge conjugate of the right-handed neutrinos have the same
charge as the down quarks the Majorana mass matrix will be constrained by
this charge assignment. For example, for a zero $\Sigma $ charge for the
Higgs generating this mass term, $\nu _{i}^{R}\nu _{j}^{R}\Sigma $, the
Majorana mass matrix has the same form as eq.(\ref{mm}) although with a
different expansion parameter. Moreover, due to the above charge
assignments, the Dirac neutrino mass matrix is the transpose of the up-quark
mass matrix.

The structure of the up-quark mass matrix will depend on the charges of the
right-handed quarks. However as these are the same with the charges of the
left-handed leptons the mass matrix will be constrained by the need to
generate large mixing for atmospheric neutrinos. Assigning the left-handed
leptons charges $y,x$ and 0, and the right-handed leptons charges $a,b$ and
0 we see that maximal (2-3) mixing requires $x=0$. \footnote{%
Acceptable solutions may also be generated for $x=\pm 1/2$ which gives large
but non-maximal mixing. We discuss an example of this in detail later.} Then:

\begin{equation}
M_{\ell }\propto \left( 
\begin{array}{ccc}
\bar{\epsilon}^{|a+y|} & \bar{\epsilon}^{|b+y|} & \bar{\epsilon}^{|y|} \\ 
\bar{\epsilon}^{|a|} & \bar{\epsilon}^{|b|} & 1 \\ 
\bar{\epsilon}^{|a|} & \bar{\epsilon}^{|b|} & 1
\end{array}
\right)
\end{equation}
The lepton eigenvalues are of order $1,\bar{\epsilon}^{|b|}$ and $max(\bar{%
\epsilon}^{|a+y|},\bar{\epsilon}^{|a|}\cdot \bar{\epsilon}^{|b+y|}/\bar{%
\epsilon}^{|b|})$. Fitting $m_{e}/m_{\mu }$ constrains the combined charges $%
a,y$ and $b$ for a given choice of the expansion parameter.

Now, we are ready to go to the up-quark mass matrix: Its form, for $x=0$,
is: 
\begin{equation}
M_{up}\propto \left( 
\begin{array}{ccc}
{\epsilon }^{|-4+y|} & {\epsilon }^{4} & {\epsilon }^{4} \\ 
{\epsilon }^{|1+y|} & {\epsilon } & \epsilon \\ 
{\epsilon }^{|y|} & 1 & 1
\end{array}
\right)
\end{equation}

In order to obtain the correct value for $m_{c}/m$ we need to make the
assignment $\epsilon =\bar{\epsilon}^{4}=0.23^{4}$ (where $\epsilon $ is the
up-matrix expansion parameter and $\bar{\epsilon}$ the down-quark and the
charged lepton one). Note that this is a direct outcome of the requirement
to obtain maximal (2-3) charged lepton mixing, which constrained $x$ to zero.%
{\em \ }Finally{\em \ }we can chose $y$ so as to get the correct $%
m_{u}/m_{c} $ ratio. An obvious choice is to take $y=2$ (remember that $%
\epsilon $ is now $\bar{\epsilon}^{4}$).

Turning to the implications for the mixing angles we see that the
contribution from the up quark sector to $V_{cb}$ is very small. Thus from
eq.(\ref{mm}) we see that the expectation is that $V_{cb}\simeq \sqrt{%
m_{s}/m_{b}.}$ This is too large and requires a very small coefficient in
the (2,3) entry of eq.(\ref{mm}). For this reason, we consider that this
model seems less promising in the framework of a single $U(1)$ symmetry%
\footnote{%
Note that in realistic models with more $U(1)$ groups coming from the
string, solutions have been found in \cite{ELLN} .}.

\subsection{$SU(3)_{c}\otimes SU(3)_{L}\otimes SU(3)_{R}\otimes U(1)\label%
{su33}$}

This is a particular GUT group which readily emerges from an underlying
string theory with an intrinsic $E(6)$ symmetry. In it a single family of
quarks and leptons are accommodated in a $(3,3,1)\oplus (\bar{3},1,\bar{3}%
)\oplus (1,3,\bar{3})$ under $SU(3)_{c}\otimes SU(3)_{L}\otimes SU(3)_{R}.$
The left- and right-handed quarks belong to $(3,3,1)$ and $(\bar{3},1,\bar{3}%
)$ respectively and thus their $U(1)$ charges are not related. On the other
hand the left handed and (charge conjugate) right handed leptons belong to
the same $(1,3,\bar{3})$ representation and hence must have the same $U(1)$
charge. Thus, the lepton mass matrices have to be symmetric.

This freedom allows us to construct fully realistic mass matrices. Let us
start from the lepton mass matrices. Taking the charges 
\begin{eqnarray}
b_{i}=c_{i}=d_{i} &=&(-\frac{7}{2},\frac{1}{2},0)  \nonumber \\
b_{i}=c_{i}=d_{i} &=&(\frac{5}{2},\frac{1}{2},0)  \nonumber \\
b_{i}=c_{i}=d_{i} &=&(3,0,0)
\end{eqnarray}
leads to the three possible charged lepton matrices 
\begin{equation}
\frac{M_{\ell }}{m_{\tau }}=\left( 
\begin{array}{ccc}
\bar{\epsilon}^{7} & \bar{\epsilon}^{3} & \bar{\epsilon}^{7/2} \\ 
\bar{\epsilon}^{3} & \bar{\epsilon} & \bar{\epsilon}^{1/2} \\ 
\bar{\epsilon}^{7/2} & \bar{\epsilon}^{1/2} & 1
\end{array}
\right) ,\frac{M_{\ell }}{m_{\tau }}=\left( 
\begin{array}{ccc}
\bar{\epsilon}^{5} & \bar{\epsilon}^{3} & \bar{\epsilon}^{5/2} \\ 
\bar{\epsilon}^{3} & \bar{\epsilon} & \bar{\epsilon}^{1/2} \\ 
\bar{\epsilon}^{5/2} & \bar{\epsilon}^{1/2} & 1
\end{array}
\right) ,\frac{M_{\ell }}{m_{\tau }}=\left( 
\begin{array}{ccc}
\bar{\epsilon}^{6} & \bar{\epsilon}^{3} & \bar{\epsilon}^{3} \\ 
\bar{\epsilon}^{3} & 1 & 1 \\ 
\bar{\epsilon}^{3} & 1 & 1
\end{array}
\right) ,  \label{lmm}
\end{equation}
We see that the third matrix leads to maximal mixing, however it requires an
accurate cancellation in the (2,3) sector in order to get the correct $%
m_{\mu }/m_{\tau }$. On the other hand, the other two matrices, lead to
natural lepton hierarchies and predict large but non-maximal lepton mixing.
We study this case in detail in section \ref{sec:tz}.

What about the quark mass matrices? The choice of $U(1)$ charges given by 
\begin{eqnarray}
\alpha _{i} &=&(3,2,0)  \nonumber \\
\beta _{i} &=&\gamma _{i}=(1,0,0)
\end{eqnarray}
leads to the mass matrices 
\[
\frac{M_{u}}{m_{t}}=\left( 
\begin{array}{ccc}
{\epsilon }^{4} & {\epsilon }^{3} & {\epsilon }^{3} \\ 
{\epsilon }^{3} & {\epsilon }^{2} & {\epsilon }^{2} \\ 
{\epsilon } & 1 & 1
\end{array}
\right) ,\frac{M_{down}}{m_{b}}=\left( 
\begin{array}{ccc}
\bar{\epsilon}^{4} & \bar{\epsilon}^{3} & \bar{\epsilon}^{3} \\ 
\bar{\epsilon}^{3} & \bar{\epsilon}^{2} & \bar{\epsilon}^{2} \\ 
\bar{\epsilon} & 1 & 1
\end{array}
\right) 
\]
For $\epsilon =\bar{\epsilon}^{2}$ we obtain viable quark hierarchies. The $%
V_{CKM}$ mixing is dominated by the contribution from the down quark sector.
However the structure of charges chosen here means that $V_{cb}\simeq
m_{s}/m_{b}.$ This is in good agreement with the measured value.

The structure of the neutrino mass matrices is fixed because the left and
right handed neutrino charges are determined because they belong to the same 
$(1,3,\bar{3})$ representation as the charged leptons. Both the Dirac and
Majorana mass matrices have a symmetric form with the Dirac mass matrix of
the same form as the charged leptons but with a different expansion
parameter. This case is discussed in detail in the next section where we
consider symmetric mass matrices in general. The first two forms of eq.(\ref
{lmm}) lead to large mixing in the atmospheric neutrino sector and generate
solar neutrino oscillation with parameters in the small mixing angle range
of eq.(\ref{atmos}).

\subsection{Left-Right symmetric models.}

Another gauge structure that has been widely explored is one which is
left-right symmetric \cite{mohapatra}. The simplest possibility is $%
SU(3)_{c}\otimes SU(2)_{L}\otimes SU(2)_{R}\otimes U(1)_{B-L}\otimes U(1)$
with a discrete $Z_{2}$ symmetry interchanging the two $SU(2)$ factors and
their associated quark and lepton representations. This is readily
generalised to larger groups. For example, the case considered above $%
SU(3)_{c}\otimes SU(3)_{L}\otimes SU(3)_{R}\otimes U(1)$ may similarly be
rendered left-right symmetric through a $Z_{2}$ symmetry interchanging the
two $SU(3)$ factors. Such a structure is found in the three generation
string theories resulting from compactifying on a specific Calabi-Yau
manifold \cite{green}. A partial unification based on $SU(2)_{L}\otimes
SU(2)_{R}\otimes SU(4)_{C}$ was proposed by Pati and Salam \cite{pati}.

In these models the U(1) family charges are strongly constrained because the 
$Z_{2}$ symmetry requires that the $U(1)$ charges of the left- and right-
handed fields be the same. As a result the mass matrices will be symmetric.
The left- right- symmetry together with the $SU(2)_{L}$ symmetry requires
that the left- and right-handed components of up and down quarks of each
generation should have the same charge and the left- and right-handed
components of the charged leptons and neutrinos of each generation should
have also have the same charge. This means that only six $U(1)$ charges need
to be specified to completely fix the model. Given the interest in left-
right- symmetric models and the highly constrained nature of the $U(1)$ we
think it of some interest to explore this possibility in some detail. As we
shall discuss the fully left- right- symmetric models with symmetric quark,
lepton and neutrino masses, are in remarkably good agreement with the
measured values of masses and mixing angles, with texture zeros leading to
definite relations between masses and mixing angles in good agreement with
experiment. However as noted above, particularly in Section \ref{su33}, it
may be that only a sub-sector has a symmetric mass matrix. To deal with all
these cases in the next Section we turn to a detailed discussion of the
implications of an Abelian family symmetry leading to symmetric mass
matrices.

\section{Symmetric textures and neutrino masses \label{sec:tz}}

\subsection{Quark Masses}

\bigskip 
\begin{table}[tbp] \centering%
%
\begin{tabular}{|l|}
\hline
{\small 
\begin{tabular}{|c|cccccccc|}
\hline
& $Q_{i}$ & $u_{i}^{c}$ & $d_{i}^{c}$ & $L_{i}$ & $e_{i}^{c}$ & $\nu
_{i}^{c} $ & $H_{2}$ & $H_{1}$ \\ \hline
$U(1)_{FD}$ & $\alpha _{i}$ & $\alpha _{i}$ & $\alpha _{i}$ & $b_{i}$ & $%
b_{i}$ & $b_{i}$ & $-2\alpha _{3}$ & $-2\alpha _{3}$ \\ \hline
\end{tabular}
} \\ \hline
\end{tabular}
\label{table:2a} \caption{Symmetric $U(1)_{FD}$ charges\label{table:1a}}%
\end{table}%
%

Here we consider in more detail the implications of the symmetric charge
assignments discussed in the last section. Although we are primarily
interested in neutrino masses, in order to \ answer the question whether the
neutrino masses and mixings fit into the pattern of quark and lepton masses,
it is necessary to discuss the latter first. We start with an Abelian family
symmetry with the most general symmetric charge assignments given in Table 
\ref{table:1a}. Following the discussion of Section \ref{sec:abelian} we
find mass matrices of the form

\begin{eqnarray}
\frac{M_u}{m_t}\approx \left( 
\begin{array}{ccc}
h_{1 1}\rho_{11 }\epsilon_a^{\mid 2+6a \mid } & h_{1 2}\rho_{12
}\epsilon_b^{\mid 3a \mid } & h_{1 3}\rho_{13 }\epsilon_a^{\mid 1+3a\mid }
\\ 
h_{2 1}\rho_{21 }\epsilon_b^{\mid 3a \mid } & h_{2 2}\rho_{22 }\epsilon^{ 2 }
& h_{2 3}\rho_{23 }\epsilon^{ 1 } \\ 
h_{3 1}\rho_{31 }\epsilon_a^{\mid 1+3a \mid } & h_{3 2}\rho_{32 }\epsilon^{1
} & h_{3 3}
\end{array}
\right)  \label{eq:mu0}
\end{eqnarray}

\begin{equation}
\frac{M_{d}}{m_{b}}\approx \left( 
\begin{array}{ccc}
k_{11}\sigma _{11}\bar{\epsilon _{a}}^{\mid 2+6a\mid } & k_{12}\sigma _{12}%
\bar{\epsilon _{b}}^{\mid 3a\mid } & k_{13}\sigma _{13}\bar{\epsilon _{a}}%
^{\mid 1+3a\mid } \\ 
k_{21}\sigma _{21}\bar{\epsilon _{b}}^{\mid 3a\mid } & k_{22}\sigma _{22}%
\bar{\epsilon}^{2} & k_{23}\sigma _{23}\bar{\epsilon}^{1} \\ 
k_{31}\sigma _{31}\bar{\epsilon _{a}}^{\mid 1+3a\mid } & k_{32}\sigma _{32}%
\bar{\epsilon}^{1} & k_{33}
\end{array}
\right)  \label{eq:massu}
\end{equation}
where $\bar{\epsilon}=(\frac{<\theta >}{M_{1}})^{|\alpha _{2}-\alpha _{1}|}$%
, $\epsilon =(\frac{<\theta >}{M_{2}})^{|\alpha _{2}-\alpha _{1}|}$, and $%
a=(2\alpha _{1}-\alpha _{2}-\alpha _{3})/3(\alpha _{2}-\alpha _{1})$, $%
h_{ij},\;k_{ij}$ are Yukawa couplings all assumed to be of $O(1)$ and $\rho
,\;\sigma $ are related to Yukawa couplings in the Higgs sector (again we
expect them to be $O(1)$) and describe Higgs or quark mixing in the way
discussed below.

It is straightforward now to see how texture zeros occur. For $-3a>1$ $%
\epsilon _{a}=\epsilon _{b}=\epsilon $ and $\bar{\epsilon}_{a}=\bar{\epsilon}%
_{b}=\bar{\epsilon}$. In this case it is easy to check that there are {\it no%
} texture zeros because all matrix elements contribute at leading order to
the masses and mixing angles. For $1>-3a>0$, $\epsilon _{a},\;\bar{\epsilon}%
_{a}$ change and are given by $\bar{\epsilon _{a}}=(\frac{<\bar{\theta}>}{%
M_{1}})^{|\alpha _{2}-\alpha _{1}|}$, $\epsilon _{a}=(\frac{<\bar{\theta}>}{%
M_{2}})^{|\alpha _{2}-\alpha _{1}|}$. In this case texture zeros in the
(1,1) and (1,3) positions {\it automatically} appear for small $<\bar{\theta}%
>/M_{i}$. However the (1,2) matrix element is too large (cf. eqs. (\ref
{eq:mu0}),(\ref{eq:massu}). For $a>0$ however $\bar{\epsilon}_{a,b}=(\frac{<%
\bar{\theta}>}{M_{1}})^{|\alpha _{2}-\alpha _{1}|}$, $\epsilon _{a,b}=(\frac{%
<\bar{\theta}>}{M_{2}})^{|\alpha _{2}-\alpha _{1}|}$, the texture zeros in
the (1,1) and (1,3) positions persist, and the (1,2) matrix element can be
of the correct magnitude.

Thus we see that the simplest possibility of an additional $U(1)$ gauge
family symmetry {\it requires} texture zeros in the phenomenologically
desirable positions for a large range of the single relevant free parameter,
a. In addition it generates structure for the other matrix elements which
can duplicate the required hierarchical structure of masses and mixing
angles. To illustrate the mechanism we consider Froggatt-Nielsen mixing in
the Higgs sector masses along the lines mapped out in Section \ref{sec:2x2}.
After mixing the light Higgs states are given by $H_{33}^{2}+\sum \rho
_{ij}H_{ij}^{2}\epsilon _{a,\;,b}^{n_{ij}}$ and $H_{33}^{1}+\sum \sigma
_{ij}H_{ij}^{1}\bar{\epsilon}_{a,\;,b}^{n_{ij}}$ where the powers $n_{ij}$
are those appearing in eq.(\ref{eq:massu}) and $\rho ,\;\sigma $ are related
to Yukawa couplings in the Higgs sector. Similarly mixing in the quark
sector can also generate the elements of eq.(\ref{eq:massu}).

As discussed above, for $a>0$, there are two approximate texture zeros in
the (1,1) and (1,3), (3,1) positions. These give rise to excellent
predictions for two combinations of the CKM matrix. The magnitude of the
remaining matrix elements is sensitive to the magnitude of a and the values
of the expansion parameters. Then choosing $a=1$ the remaining non-zero
entries have magnitude in excellent agreement with the measured values. From
eqs. (\ref{eq:mu0}) and (\ref{eq:massu}), we see that to a good
approximation we have the relation \cite{IR}

\begin{equation}
\epsilon =\bar{\epsilon}^{2}  \label{eq:eps}
\end{equation}
corresponding to the choice $M_{2}>M_{1}$.

Such a choice gives an excellent description of quark masses and mixing
angles, after allowing for the unknown coefficients of $O(1).$ It is
instructive to determine the magnitude needed for these coefficients to fit
all measured masses and mixings. This is clearly an under-determined
problem, so the best we can do is to give an illustrative example. The choice

\[
\frac{M_{u}}{m_{t}}\approx \left( 
\begin{array}{ccc}
0 & \epsilon ^{3} & 0 \\ 
\epsilon ^{3} & i\epsilon ^{2} & \epsilon \\ 
0 & \epsilon & 1
\end{array}
\right) 
\]

\begin{equation}
\frac{M_{d}}{m_{b}}\approx \left( 
\begin{array}{ccc}
0 & \overline{\epsilon }^{3} & 0 \\ 
\overline{\epsilon }^{3} & \overline{\epsilon }^{2} & \frac{1}{2}\overline{%
\epsilon } \\ 
0 & \frac{1}{2}\overline{\epsilon } & 1
\end{array}
\right)
\end{equation}
generates the correct masses and mixing angles with $\epsilon =0.05,$ $%
\overline{\epsilon }=0.18.$ This requires a coefficient $1/2$ in the $(2,3)$
entry, necessary to give the value of $V_{cb}=0.04.$ The latter is
anomalously small due to a cancellation between the contributions of the up
and the down quark sectors. Such a cancellation may be expected as there is
an approximate $SU(2)_{R}$ symmetry in the magnitude of the matrix elements
following from the very symmetric choice of family charges and in the limit $%
SU(2)_{R}$ is exact $V_{cb}$ vanishes.

\subsection{Lepton Masses}

Now that we have a theory of quark masses it is possible systematically to
address the original question whether the large mixing angle found in the
neutrino sector is consistent with this theory or whether it requires
completely new structure \cite{DLLRS,LLR,ver}.

\subsubsection{Charged leptons}

The charged lepton masses and mixings are determined in a similar way to
that of the quarks. Requiring $m_{b}=m_{\tau }$ at unification, sets $\alpha
_{3}=b_{3}$ and then the charged lepton mass matrix is 
\begin{equation}
\frac{M_{\ell }}{m_{\tau }}\approx \left( 
\begin{array}{ccc}
\bar{\epsilon}^{\mid 2+6a-2b\mid } & \bar{\epsilon}^{\mid 3a\mid } & \bar{%
\epsilon}^{\mid 1+3a-b\mid } \\ 
\bar{\epsilon}^{\mid 3a\mid } & \bar{\epsilon}^{\mid 2(1-b)\mid } & \bar{%
\epsilon}^{\mid 1-b\mid } \\ 
\bar{\epsilon}^{\mid 1+3a-b\mid } & \bar{\epsilon}^{\mid 1-b\mid } & 1
\end{array}
\right)  \label{chargedlepton}
\end{equation}
where $b=(\alpha _{2}-b_{2})/(\alpha _{2}-\alpha _{3})$. A solution with $%
b=0 $ leads to lepton hierarchies similar to those for the quarks. However
in this case the expectation is that $m_{s}\simeq m_{\mu }$ and $m_{d}\simeq
m_{e}$ at the unification scale, in conflict with experiment. In this case
we must rely on large coefficients to generate an acceptable mass matrix
structure. If instead we require that the explanation of the mass structure
of the light generations results from the choice of lepton family charges,
i.e. through the choice of $b$, one is led to take\ $\beta \equiv 1-b=\pm
1/2 $\footnote{{\small In some cases, the textures with half-integer $b$
have been simplified, by imposing a residual $Z_{2}$ discrete gauge symmetry
after the $U(1)$ breaking, by which the electron and muon fields transform
by $(-1)$. Then, entries raised in a half-integer power vanish. However this
is not a necessary condition: in general, half-integer entries remain
present at low energies and may have interesting phenomenological
implications.}}.

\ Let us first comment on the case with $\beta =-1/2$. This gives the lepton
texture \cite{IR} 
\begin{equation}
\frac{M_{\ell }}{m_{\tau }}=\left( \ 
\begin{array}{ccc}
\bar{\epsilon}^{5} & \bar{\epsilon}^{3} & \bar{\epsilon}^{5/2} \\ 
\bar{\epsilon}^{3} & \bar{\epsilon} & \bar{\epsilon}^{1/2} \\ 
\bar{\epsilon}^{5/2} & \bar{\epsilon}^{1/2} & 1
\end{array}
\right)  \label{b-1/2}
\end{equation}
On the other hand $\beta =1/2$, leads to \cite{DLLRS,LLR} 
\begin{equation}
\frac{M_{\ell }}{m_{\tau }}=\left( \ 
\begin{array}{ccc}
\bar{\epsilon}^{7} & \bar{\epsilon}^{3} & \bar{\epsilon}^{7/2} \\ 
\bar{\epsilon}^{3} & \bar{\epsilon} & \bar{\epsilon}^{1/2} \\ 
\bar{\epsilon}^{7/2} & \bar{\epsilon}^{1/2} & 1
\end{array}
\right)  \label{b1/2}
\end{equation}
As we see, both types of textures can give correct predictions for lepton
masses (note that when estimating the lightest eigenvalue we have not
allowed for cancellations in eq.(\ref{b-1/2}) between the contributions from
the (1,1), (2,2) and (1,2), (2,1) elements. Due to the Yukawa couplings of $%
O(1)$ we do not expect such a cancellation to occur.).

\subsubsection{Neutrinos}

We may now determine the predictions of the flavour symmetry for neutrino
masses. Due to the see-saw mechanism we generate quite naturally light
neutrino masses. In this framework, the light Majorana neutrino masses are
given by the generalisation of eq.(\ref{eigenvalues}) 
\begin{equation}
m_{eff}=M_{\nu }^{D}\cdot (M_{\nu _{R}}^{M})^{-1}\cdot (M_{\nu }^{D})^{T}
\label{eq:meff}
\end{equation}
where $M_{\nu }^{D}$ and $M_{\nu _{R}}^{M}$ are the $3\times 3$ Dirac and
Majorana mass matrices respectively.

How do we determine these mass matrices? The Dirac mass matrix is actually
fixed by the symmetries of the model. Indeed, $SU(2)_{L}$ fixes the $%
U(1)_{FD}$ charge of the left-handed neutrino states to be the same as the
charged leptons, and then the left-right symmetry fixes the charges of the
right-handed neutrinos as given in Table \ref{table:2a}. Thus the neutrino
Dirac mass is given by 
\begin{equation}
M_{\nu }^{D}\propto \left( 
\begin{array}{ccc}
{\epsilon }^{\mid 2+6a-2b\mid } & {\epsilon }^{\mid 3a\mid } & {\epsilon }%
^{\mid 1+3a-b\mid } \\ 
{\epsilon }^{\mid 3a\mid } & {\epsilon }^{\mid 2(1-b)\mid } & {\epsilon }%
^{\mid 1-b\mid } \\ 
{\epsilon }^{\mid 1+3a-b\mid } & {\epsilon }^{\mid 1-b\mid } & 1
\end{array}
\right)  \label{eq:nud}
\end{equation}
In unified or partially unified models the large mass scale associated with
Froggatt-Nielsen mixing is the same as the one for the up-quarks and so in
eq.(\ref{eq:nud}) we have used the up quark expansion parameter.

We turn now to the Majorana masses for the right-handed neutrinos. Such
masses arise from terms of the form $\nu _{R}\nu _{R}\Sigma $ where $\Sigma $
is a $SU(3)\otimes SU(2)\otimes U(1)$ invariant Higgs scalar field with $%
I_{W}=0$ and $\nu _{R}$ is a right-handed neutrino. Since we do not know the
charge of $\Sigma $, we have to consider all possible choices. This allows
us to ``rotate'' the larger coupling to any of the entries of the heavy
Majorana mass matrix, generating a discrete spectrum of possible forms \cite
{DLLRS,LLR}. For example, if the $\Sigma $ charge is the same as that of the 
$H_{1,2}$ doublet Higgs charges, the larger element of $M_{\nu }$ will be in
the (3,3) entry. The rest of the terms will be generated as before through
the $U(1)_{FD}$ breaking by $<\theta >$ and $<\bar{\theta}>$.

Among the cases that naturally generate the correct lepton hierarchies those
that are also of interest for Super-Kamiokande are mainly $\beta \equiv
1-b=\pm 1/2$ which lead to large (2-3) lepton mixing \footnote{{\small We
discussed the first case in \cite{LLR}.}}. Restricting the discussion to
these cases the general forms for the heavy Majorana mass textures (allowing
for the various choices of the $\Sigma $ charge), appear in Table \ref
{table:maj}. The form of $m_{eff}$, its eigenvalues and the mixing matrices
for $\beta =1/2$, are presented in Table \ref{table:meff2} and for $\beta
=-1/2$ in Table \ref{table:meff}. Here, $\tilde{m}_{eff}=m_{\nu
,diag}^{D}\cdot R_{D}^{T}\cdot (M_{\nu }^{R})^{-1}\cdot R_{D}\cdot m_{\nu
,diag}^{D}$, where $R_{D}$ is the neutrino Dirac mixing matrix. This
combination has been chosen because it contains all the information
necessary to determine the mass eigenvalues and also exhibits the
contribution to the mixing angles that is sensitive to the mixing in the
Majorana mass matrix. The full $\ $effective Majorana mass matrix is then
given by $m_{eff}=R_{D}.\tilde{m}_{eff}.R_{D}^{T}.$ A word of caution is in
order here. The mass hierarchies are quite sensitive to the order unity
coefficients that are not be predicted by the $U(1)$ symmetry due to the
fact that the inverse of the Majorana mass matrix must be taken and the
product of several mass matrices are involved. Indeed a small difference in
a coefficient in the neutrino Dirac mass matrix, may lead to a large
difference in the eigenvalues of $m_{eff}$. Thus the estimate given in
Tables \ref{table:meff2} and \ref{table:meff} for $m_{eff}$ should be viewed
as only a rough estimate. On the other hand the sensitivity of the mixing
angles to the $O(1)$ coefficients is much less because, as discussed above,
it largely comes from the Dirac mass matrix alone. Even with this cautionary
word we see from Tables \ref{table:meff2} and\ \ref{table:meff} we see that
in all cases large mass hierarchies between the neutrino masses are expected
to arise. The lightest neutrino eigenvalue is very suppressed compared to
the other two.

\begin{table}[tbp]
{\small \centering
\begin{tabular}{|c|c|}
\hline
$\left ( 
\begin{array}{ccc}
\bar\eps^{2\mid 3\alpha + \beta\mid } & \bar\eps^{3\mid\alpha\mid} & \bar\eps%
^{\mid 3\alpha + \beta\mid } \\ 
\bar\eps^{\mid 3\alpha \mid } & \bar\eps^{2\mid \beta \mid} & \bar\eps^{\mid
\beta \mid} \\ 
\bar\eps^{\mid 3\alpha + \beta\mid } & \bar\eps^{\mid \beta \mid} & 1
\end{array}
\right)$ & $\left ( 
\begin{array}{ccc}
\bar\eps^{3\mid 2\alpha + \beta\mid } & \bar\eps^{\mid 3\alpha + \beta\mid}
& \bar\eps^{\mid 3\alpha + 2\beta\mid } \\ 
\bar\eps^{\mid 3\alpha + \beta\mid } & \bar\eps^{\mid \beta \mid} & 1 \\ 
\bar\eps^{\mid 3\alpha + 2\beta\mid } & 1 & \bar\eps^{\mid \beta \mid}
\end{array}
\right)$ \\ \hline
$\left ( 
\begin{array}{ccc}
\bar\eps^{2\mid 3\alpha + 2\beta\mid } & \bar\eps^{\mid 3\alpha + 2\beta\mid}
& \bar\eps^{3\mid \alpha + \beta\mid } \\ 
\bar\eps^{\mid 3\alpha + 2\beta\mid } & 1 & \bar\eps^{\mid \beta \mid} \\ 
\bar\eps^{3\mid \alpha + \beta\mid } & \bar\eps^{\mid \beta \mid} & \bar\eps%
^{2\mid \beta \mid}
\end{array}
\right)$ & $\left ( 
\begin{array}{ccc}
\bar\eps^{\mid 3\alpha + \beta\mid } & \bar\eps^{\mid \beta\mid} & 1 \\ 
\bar\eps^{ \mid\beta\mid } & \bar\eps^{3\mid \alpha + \beta\mid } & \bar\eps%
^{3\mid \alpha + 2\beta\mid } \\ 
1 & \bar\eps^{3\mid \alpha + 2\beta\mid } & \bar\eps^{\mid 3\alpha + \beta
\mid}
\end{array}
\right)$ \\ \hline
$\left ( 
\begin{array}{ccc}
1 & \bar\eps^{\mid 3\alpha + 2\beta\mid } & \bar\eps^{\mid 3\alpha +
\beta\mid} \\ 
\bar\eps^{\mid 3\alpha + 2\beta\mid } & \bar\eps^{2\mid 3\alpha + 2\beta\mid
} & \bar\eps^{3\mid 2\alpha + \beta\mid } \\ 
\bar\eps^{\mid 3\alpha + \beta\mid} & \bar\eps^{3\mid 2\alpha + \beta\mid }
& \bar\eps^{2\mid 3\alpha + \beta \mid}
\end{array}
\right)$ & $\left( 
\begin{array}{ccc}
\bar\eps^{\mid 3\alpha + 2\beta\mid } & 1 & \bar\eps^{\mid\beta\mid } \\ 
1 & \bar\eps^{\mid 3\alpha + 2\beta\mid } & \bar\eps^{\mid 3\alpha +
\beta\mid } \\ 
\bar\eps^{\mid\beta\mid } & \bar\eps^{\mid 3\alpha + \beta\mid} & \bar\eps%
^{\mid 3\alpha\mid}
\end{array}
\right)$ \\ \hline
\end{tabular}
}
\caption{General forms of heavy Majorana mass matrix textures. Interesting
textures arise for $\protect\alpha =1,\protect\beta \pm 1/2$.}
\label{table:maj}
\end{table}

\begin{table}[tbp]
{\small \centering
\begin{tabular}{|c|c|c|c|c|}
\hline
& $\tilde{m}^{eff}_{\nu}$ & $m^{eff,Diag}_{\nu}$ & $m^{eff}_{\nu}$ & $%
R_{\nu}^{eff}$ \\ \hline
1 & $\left( 
\begin{array}{ccc}
e^{30} & e^{18} & e^{15} \\ 
e^{18} & e^{10} & e^5 \\ 
e^{15} & e^5 & 1
\end{array}
\right)$ & $\left( 
\begin{array}{ccc}
e^{26} &  &  \\ 
& e^{10} &  \\ 
&  & 1
\end{array}
\right)$ & $\left( 
\begin{array}{ccc}
e^{26} & e^{15} & e^{13} \\ 
e^{15} & e^{4} & e^2 \\ 
e^{13} & e^2 & 1
\end{array}
\right)$ & $\left( 
\begin{array}{ccc}
1 & e^{11} & -e^{13} \\ 
-e^{11} & 1 & e^2 \\ 
e^{13} & -e^2 & 1
\end{array}
\right)$ \\ \hline
2 & $\left( 
\begin{array}{ccc}
e^{25} & e^{17} & e^{12} \\ 
e^{17} & e^{9} & e^4 \\ 
e^{12} & e^4 & e^{-1}
\end{array}
\right)$ & $\left( 
\begin{array}{ccc}
e^{25} &  &  \\ 
& e^{9} &  \\ 
&  & e^{-1}
\end{array}
\right)$ & $\left( 
\begin{array}{ccc}
e^{25} & e^{14} & e^{12} \\ 
e^{14} & e^{3} & e \\ 
e^{12} & e & e^{-1}
\end{array}
\right)$ & $\left( 
\begin{array}{ccc}
1 & e^{11} & -e^{13} \\ 
-e^{11} & 1 & e^2 \\ 
e^{13} & -e^2 & 1
\end{array}
\right)$ \\ \hline
3 & $\left( 
\begin{array}{ccc}
e^{24} & e^{16} & e^{11} \\ 
e^{16} & e^{8} & e^{3} \\ 
e^{11} & e^{3} & e^{-2}
\end{array}
\right)$ & $\left( 
\begin{array}{ccc}
e^{24} &  &  \\ 
& e^{8} &  \\ 
&  & e^{-2}
\end{array}
\right)$ & $\left( 
\begin{array}{ccc}
e^{24} & e^{13} & e^{11} \\ 
e^{13} & e^{2} & 1 \\ 
e^{11} & 1 & e^{-2}
\end{array}
\right)$ & $\left( 
\begin{array}{ccc}
1 & e^{11} & -e^{13} \\ 
-e^{11} & 1 & e^2 \\ 
e^{13} & -e^2 & 1
\end{array}
\right)$ \\ \hline
4 & $\left( 
\begin{array}{ccc}
e^{47} & e^{23} & e^{20} \\ 
e^{23} & e^{-1} & e^{-4} \\ 
e^{20} & e^{-4} & e^{-7}
\end{array}
\right)$ & $\left( 
\begin{array}{ccc}
e^{33} &  &  \\ 
& e^{13} &  \\ 
&  & e^{-7}
\end{array}
\right)$ & $\left( 
\begin{array}{ccc}
e^{15} & e^{6} & e^{4} \\ 
e^{6} & e^{-3} & e^{-5} \\ 
e^{4} & e^{-5} & e^{-7}
\end{array}
\right)$ & $\left( 
\begin{array}{ccc}
1 & e^{9} & -e^{11} \\ 
-e^{9} & 1 & e^2 \\ 
e^{11} & -e^2 & 1
\end{array}
\right)$ \\ \hline
5 & $\left( 
\begin{array}{ccc}
e^{40} & e^{16} & e^{13} \\ 
e^{16} & e^{-8} & e^{-11} \\ 
e^{13} & e^{-11} & e^{-14}
\end{array}
\right)$ & $\left( 
\begin{array}{ccc}
e^{40} &  &  \\ 
& e^{-8} &  \\ 
&  & e^{-14}
\end{array}
\right)$ & $\left( 
\begin{array}{ccc}
e^{8} & e^{-1} & e^{-3} \\ 
e^{-1} & e^{-10} & e^{-12} \\ 
e^{-3} & e^{-12} & e^{-14}
\end{array}
\right)$ & $\left( 
\begin{array}{ccc}
1 & e^{9} & -e^{7} \\ 
-e^{9} & 1 & e^2 \\ 
e^{7} & -e^2 & 1
\end{array}
\right)$ \\ \hline
6 & $\left( 
\begin{array}{ccc}
e^{48} & e^{24} & e^{21} \\ 
e^{24} & e^{4} & e^{-1} \\ 
e^{21} & e^{-1} & e^{-6}
\end{array}
\right)$ & $\left( 
\begin{array}{ccc}
e^{32} &  &  \\ 
& e^{16} &  \\ 
&  & e^{-6}
\end{array}
\right)$ & $\left( 
\begin{array}{ccc}
e^{20} & e^{9} & e^{7} \\ 
e^{9} & e^{-2} & e^{-4} \\ 
e^{7} & e^{-4} & e^{-6}
\end{array}
\right)$ & $\left( 
\begin{array}{ccc}
1 & e^{11} & -e^{13} \\ 
-e^{11} & 1 & e^2 \\ 
e^{13} & -e^2 & 1
\end{array}
\right)$ \\ \hline
\end{tabular}
}
\caption{Masses and mixing angles for the light neutrino components, and for 
$b=1/2$.}
\label{table:meff2}
\end{table}

\begin{table}[tbp]
{\small \centering
\begin{tabular}{|c|c|c|c|c|}
\hline
& $\tilde{m}^{eff}_{\nu}$ & $m^{eff,Diag}_{\nu}$ & $m^{eff}_{\nu}$ & $%
R_{\nu}^{eff}$ \\ \hline
1 & $\left( 
\begin{array}{ccc}
e^{30} & e^{18} & e^{15} \\ 
e^{18} & e^{6} & e^3 \\ 
e^{15} & e^3 & 1
\end{array}
\right)$ & $\left( 
\begin{array}{ccc}
e^{30} &  &  \\ 
& e^{6} &  \\ 
&  & 1
\end{array}
\right)$ & $\left( 
\begin{array}{ccc}
e^{20} & e^{12} & e^{10} \\ 
e^{12} & e^{4} & e^2 \\ 
e^{10} & e^2 & 1
\end{array}
\right)$ & $\left( 
\begin{array}{ccc}
1 & e^{8} & -e^{10} \\ 
-e^{8} & 1 & e^2 \\ 
e^{10} & -e^2 & 1
\end{array}
\right)$ \\ \hline
2 & $\left( 
\begin{array}{ccc}
e^{31} & e^{19} & e^{16} \\ 
e^{19} & e^{7} & e^4 \\ 
e^{16} & e^4 & e
\end{array}
\right)$ & $\left( 
\begin{array}{ccc}
e^{31} &  &  \\ 
& e^{7} &  \\ 
&  & e
\end{array}
\right)$ & $\left( 
\begin{array}{ccc}
e^{21} & e^{13} & e^{11} \\ 
e^{13} & e^{5} & e^3 \\ 
e^{11} & e^3 & e
\end{array}
\right)$ & $\left( 
\begin{array}{ccc}
1 & e^{8} & -e^{10} \\ 
-e^{8} & 1 & e^2 \\ 
e^{10} & -e^2 & 1
\end{array}
\right)$ \\ \hline
3 & $\left( 
\begin{array}{ccc}
e^{36} & e^{22} & e^{17} \\ 
e^{22} & e^{8} & e^{5} \\ 
e^{17} & e^{5} & e^{2}
\end{array}
\right)$ & $\left( 
\begin{array}{ccc}
e^{32} &  &  \\ 
& e^{8} &  \\ 
&  & e^{2}
\end{array}
\right)$ & $\left( 
\begin{array}{ccc}
e^{22} & e^{14} & e^{12} \\ 
e^{14} & e^{6} & e^4 \\ 
e^{12} & e^4 & e^2
\end{array}
\right)$ & $\left( 
\begin{array}{ccc}
1 & e^{8} & -e^{10} \\ 
-e^{8} & 1 & e^2 \\ 
e^{10} & -e^2 & 1
\end{array}
\right)$ \\ \hline
4 & $\left( 
\begin{array}{ccc}
e^{39} & e^{23} & e^{20} \\ 
e^{23} & e^{7} & e^{4} \\ 
e^{20} & e^{4} & e
\end{array}
\right)$ & $\left( 
\begin{array}{ccc}
e^{35} &  &  \\ 
& e^{11} &  \\ 
&  & e
\end{array}
\right)$ & $\left( 
\begin{array}{ccc}
e^{21} & e^{13} & e^{11} \\ 
e^{13} & e^{5} & e^{3} \\ 
e^{11} & e^{3} & e
\end{array}
\right)$ & $\left( 
\begin{array}{ccc}
1 & e^{8} & -e^{10} \\ 
-e^{8} & 1 & e^2 \\ 
e^{10} & -e^2 & 1
\end{array}
\right)$ \\ \hline
5 & $\left( 
\begin{array}{ccc}
e^{40} & e^{20} & e^{15} \\ 
e^{20} & 1 & e^{-5} \\ 
e^{15} & e^{-5} & e^{-10}
\end{array}
\right)$ & $\left( 
\begin{array}{ccc}
e^{40} &  &  \\ 
& 1 &  \\ 
&  & e^{-10}
\end{array}
\right)$ & $\left( 
\begin{array}{ccc}
e^{10} & e^2 & 1 \\ 
e^{2} & e^{-6} & e^{-8} \\ 
1 & e^{-8} & e^{-10}
\end{array}
\right)$ & $\left( 
\begin{array}{ccc}
1 & e^{8} & -e^{10} \\ 
-e^{8} & 1 & e^2 \\ 
e^{10} & -e^2 & 1
\end{array}
\right)$ \\ \hline
6 & $\left( 
\begin{array}{ccc}
e^{44} & e^{24} & e^{19} \\ 
e^{24} & e^{4} & e^{-1} \\ 
e^{19} & e^{-1} & e^{-6}
\end{array}
\right)$ & $\left( 
\begin{array}{ccc}
e^{36} &  &  \\ 
& e^{12} &  \\ 
&  & e^{-6}
\end{array}
\right)$ & $\left( 
\begin{array}{ccc}
e^{14} & e^{6} & e^{4} \\ 
e^{6} & e^{-2} & e^{-4} \\ 
e^{4} & e^{-4} & e^{-6}
\end{array}
\right)$ & $\left( 
\begin{array}{ccc}
1 & e^{8} & -e^{10} \\ 
-e^{8} & 1 & e^2 \\ 
e^{10} & -e^2 & 1
\end{array}
\right)$ \\ \hline
\end{tabular}
}
\caption{Masses and mixing angles for the light neutrino components, and for 
$b=-1/2$.}
\label{table:meff}
\end{table}

\subsubsection{Neutrino Masses}

Given the results of Tables \ref{table:meff2} and\ \ref{table:meff} we may
now determine the expectation for the magnitude of the neutrino masses. As
discussed in Section \ref{sec:neutrinomass}, the double see-saw gives the
largest mass in the range required to explain the atmospheric neutrino
oscillation. Given that the Abelian family symmetry fixes the ratio of
masses can we simultaneously accommodate atmospheric and solar neutrino
oscillations? Remarkably this proves to be easy for 4 of the 12 cases of
Tables \ref{table:meff2} and\ \ref{table:meff} lead to the second heaviest
neutrino in the range needed to explain solar neutrino oscillations. In case
5 of Table \ref{table:meff2} and cases 1, 2 and 3 of Table \ref{table:meff}
we see that the ratio of the two heaviest eigenvalue is $O(e^{6})\simeq
10^{-2}.$ Thus if the heaviest neutrino has mass $0.1eV$, consistent with
atmospheric neutrino oscillation, the next neutrino will have mass $%
O(10^{-3} ~{\rm eV}).$ Given the uncertainties due to the coefficients of $%
O(1),$ this is certainly in the range needed to generate solar neutrino
oscillations via the small angle MSW solution, eq.(\ref{smallmsw}).

\subsubsection{Lepton Mixing Angles}

What about the mixing angles associated with atmospheric and solar
oscillations? The former is governed by the $(2-3)$ mixing angle. From
Tables \ref{table:meff2} and\ \ref{table:meff} we see the contribution to
this mixing angle from the neutrino sector is always of $O(e^{2}=\overline{%
\epsilon }).$ Further the contribution to this from $\tilde{m}_{eff}$ is
only of $O(e^{5})$ and so the dominant contribution is from the rotation $%
R_{D}$ needed to diagonalise the Dirac neutrino mass matrix. This is
determined entirely by the left-handed neutrino charge (which is the same as
the associated charged lepton). The contribution to the $(2,3)$ mixing angle
from the charged lepton sector may be read from eq.(\ref{b-1/2}) and eq.(\ref
{b1/2}). It is of $O(e=\sqrt{\overline{\epsilon }})$ and thus is larger than
the contribution from the neutrino matrix. Whether these two contributions
are of the same sign depends on the phases of the mass matrix elements which
is not determined by the Abelian family symmetry. In the case that the two
sources of mixing act constructively, a $(2-3)$\ mixing with $sin\theta $ up
to $\sqrt{\overline{\epsilon }}+\overline{\epsilon }\approx 0.7$ is
obtained. Given the uncertainties of the coefficients of $O(1)$ one must
conclude this is\ cconsistent with the present range observed in atmospheric
neutrino oscillation, eq.(\ref{atmos})! The conclusion is that the mixing
angles in the lepton sector are much larger than in the quark sector results
from the need to choose different family charges to account for the relative
enhancement observed for the ratio $m_{\mu }/m_{s}$ compared to $m_{\tau
}/mb.$ It is this fact that allows the large mixing observed in atmospheric
neutrino oscillation to be accommodated with the quark masses and mixings
within the context of a very simple Abelian family symmetry.

Remarkably the family symmetry also leads to an excellent prediction for the
mixing angle in the solar neutrino sector. We observed in the previous
section that four of the twelve possible structures of Tables \ref
{table:meff2} and\ \ref{table:meff} give the second heaviest neutrino in the
mass range required to explain the solar neutrino deficit. We may see from
eq.(\ref{chargedlepton}) and Tables \ref{table:meff2} and\ \ref{table:meff}
that the $(1-2)$ mixing relevant to the solar neutrino oscillations is
dominated by the mixing in the charged lepton sector and is of $O(\overline{%
\epsilon }^{2}\simeq 0.03).$ This is in excellent agreement with the range
for the small angle MSW solution, eq.(\ref{smallmsw}), which is the one
selected by the neutrino mass estimates given above.

\subsubsection{Renormalisation group stability}

Let us finally make some comments on the stability of our solutions with
respect to radiative corrections. Note that for large $tan\beta $,
renormalisation group effects tend to amplify the (2-3) mixing angle in $%
m_{eff}$. The running of the mixing angle is given by \cite{BP} 
\begin{equation}
16\pi ^{2}{\frac{d}{dt}}{\sin ^{2}2\theta _{23}}=-2\sin ^{2}2\theta
_{23}(1-\sin ^{2}2\theta _{23})(Y_{E3}^{2}-Y_{E2}^{2}){\frac{%
m_{eff}^{33}+m_{eff}^{22}}{m_{eff}^{33}-m_{eff}^{22}}}\   \label{MIXING}
\end{equation}
Due to the effect of the $\tau $ Yukawa coupling, $m_{eff}^{33}$ decreases
more rapidly than $m_{eff}^{22}$, so if at the GUT scale $%
m_{eff}^{22}<m_{eff}^{33}$, the mixing becomes larger. This can be easily
shown by semi-analytic equations \cite{ELLN}. In fact, in solutions with $%
m_{eff}^{22}$ close to $m_{eff}^{33}$, we can end up either amplifying or
destroying the mixing at the GUT scale (the later would occur if $%
m_{eff}^{22}$ becomes equal to $m_{eff}^{33}$ at an intermediate scale).
However, in the case discussed here, we get large hierarchies between $%
m_{eff}^{22}$ and $m_{eff}^{33}$ due to the splitting in the Dirac mass
matrices. This means that our solutions are stable under the RGE runs, even
for large $tan\beta $ (for small $tan\beta $ the running of $h_{\tau }$ is
so slow, that unless $m_{eff}^{22}$ and $m_{eff}^{33}$ are very close to
start with, they never become equal).

\section{Summary and Conclusions}

The measurement of neutrino oscillations interpreted as evidence for
non-zero neutrino mass has the dramatic implication that the Standard Model
in its original form is dead. However the simplest modification needed to
allow for neutrino masses, the introduction of right-handed neutrinos, is
relatively modest and has the aesthetic advantage of restoring the symmetry
between the left-handed and right-handed multiplet structure. Moreover, the
see-saw mechanism offers a very plausible explanation for the smallness of
the neutrino masses, with the heaviest neutrino quite naturally in the range
needed to explain the atmospheric neutrino oscillation in the case a double
see-saw is operative.

The Standard Model thus extended with three right-handed neutrinos is able
to generate both atmospheric and solar neutrino oscillations. Remarkably,
the neutrino masses and mixing angles needed fit quite comfortably with a
theory of quark and lepton masses based on an Abelian family symmetry with
left- right- symmetric charges, spontaneously broken at a very high mass
scale close to the gauge unification scale. In this model the reason the
mixing angles are very large in the $(2-3)$ lepton sector and less so in the 
$(2-3)$ quark sector may be traced to the fact that the {\small ratio }$%
m_{\mu }/m_{s}$ is much larger than that of $m_{\tau }/m_{b}.$ To fit this,
requires a choice of the lepton family charges which in turn gives rise to
the expectation of large mixing angles. The size of the lepton mixing in the 
$(2-3)$ sector relative to that in the quark sector (generating $V_{cb})$
may be further enhanced if in the quark sector there is a cancellation
between the rotations needed in the up and the down quark matrices, while in
the lepton sector the up and down mixings add. This is certainly possible
within the framework of an Abelian family symmetry, but is not guaranteed
because the Abelian family symmetry does not determine the relative phases
of these terms. The family symmetry {\it does} determine the order of
magnitude of the neutrino mass differences, and readily generates a mass
consistent with the small angle MSW solution to solar neutrino oscillations.
Again it is remarkable that the associated expectation for the mixing angle
is in the correct range to accommodate the small angle solution.

While the left- right- symmetric models are of particular interest because
the $U(1)$ family charges are so strongly constrained, there are further
possibilities of interest involving a combination of the family symmetry
with an extension of the Standard model gauge group. We explored various
possibilities to see if the large mixing angle observed in atmospheric
neutrino oscillation was compatible with the quark and charged lepton masses
and mixings without requiring large coefficients or cancellation between
terms unrelated by the Abelian symmetry. The most interesting possibility we
identified involved the gauge group $SU(3)^{3}.$ This has the merit that the
(small) value of $V_{cb}$ is not related to the large neutrino mixing angle.
In this the freedom to choose different left- and right- handed charges for
the down quarks allows the construction of a model with $V_{cb}\approx
m_{s}/m_{b}.$ However the left- and right-handed leptons belong to the same
representation of $SU(3)^{3}$ and this leads to the same symmetric lepton
mass matrices as in the solutions found in the fully left-right symmetric
case, thus maintaining the good prediction for large atmospheric and small
solar neutrino mixing.

We find it quite encouraging that the simplest possible family symmetry is
able to correlate so many different features of quark, charged lepton and
now neutrino masses and mixings. Due to the unknown Yukawa couplings of O(1)
it is difficult to make precise predictions (apart from those arising from
texture zeros) and this makes the scheme difficult to establish
definitively. However there are general characteristic features in the
neutrino sector that will be tested in the future. For example, although the
mixing is expected to be large in the $(2,3)$ sector, it is quite unlikely
to be maximal. Also, if the indication for neutrino masses coming from the
LSND collaboration proves correct, it will be necessary to add at least a
further sterile neutrino component. The information on neutrino masses and
mixing is very important in testing theories of fermion mass and we look
forward to the extensive new data in this area that will be forthcoming with
the new generation of detectors.

\end{document}